\documentclass[english,pra,aps,twocolumn,floatfix,tightenlines,superscriptaddress,amsmath,amssymb]{revtex4}
\usepackage[]{fontenc}
\usepackage[latin1]{inputenc}
\usepackage{verbatim}
\usepackage{graphicx}
\usepackage{physics}
\makeatletter

\def\1{1\negthickspace{\rm I}}

\usepackage{hyperref}
\hypersetup{
    colorlinks=true,
    linkcolor=cyan,
    filecolor=magenta,      
    urlcolor=blue,
    citecolor = blue,
}

\usepackage{babel}
\makeatother
\begin{document}

\title{Local entanglement and String Order Parameter in dimerized models}

  \author {Murod S. Bahovadinov}
 
\affiliation{Department of Physics, Bilkent University, 06800, Bilkent, Ankara, Turkey}
\affiliation{Department of Physics, School of Science and Technology, Nazarbayev University, 53, Kabanbay batyr Av., Astana, 010000, Republic of Kazakhstan}
\affiliation{Universit\"{a}t Bielefeld, Fakult\"{a}t f\"{u}r Physik, D-33501 Bielefeld, Germany}

 \email{murodbahovadinov@bilkent.edu.tr}
\author{O\u{g}uz G\"{u}lseren}
\affiliation{Department of Physics, Bilkent University, 06800, Bilkent, Ankara, Turkey}
 \email{gulseren@fen.bilkent.edu.tr}
 \author {J\"{u}rgen Schnack }
\affiliation{Universit\"{a}t Bielefeld, Fakult\"{a}t f\"{u}r Physik, D-33501 Bielefeld, Germany}
%\pacs{03.67.Hk, 03.67.Mn, 75.10.Pq}

\begin{abstract}
In this letter, we propose an application of String Order Parameter (SOP), commonly used in quantum spin systems, to identify symmetry-protected topological phase (SPT) in fermionic systems in the example of the dimerized fermionic chain. As a generalized form of dimerized model, we consider a one-dimensional spin-1/2 XX model with alternating spin couplings. We employ Jordan-Wigner fermionization to map this model to the spinless Su-Schrieffer-Heeger fermionic model (SSH) with generalized hopping signs. We demonstrate a phase transition between a trivial insulating phase and the Haldane phase by the exact analytical evaluation of reconstructed SOPs which are represented as determinants of Toeplitz matrices with the given generating functions. To get more insight into the topological quantum phase transition (tQPT) and microscopic correlations, we study the pairwise concurrence as a local entanglement measure of the model. We show that the first derivative of the concurrence has a non-analytic behaviour in the vicinity of the tQPT, like in the second order trivial QPTs.

\end{abstract}

\date{April 22, 2019}

\maketitle

\section{Introduction}

A detailed study of the phases of matter and  also transition between them have long been an actual problem of condensed matter physics. While Landau-Ginzburg theory can provide such understanding and explain trivial quantum phase transitions in terms of \textit{'spontaneous symmetry-breaking'}, topological phases of matter and their transitions can not be characterized in the frame of the theory.  Particularly, this is due to the lack of a local order parameter that can identify a topological phase. Consequently, exploration of these non-trivial quantum phases and their transitions has been an actively studied topic in the field of condensed matter physics.    

The oldest example of a \textit{symmetry-protected topological} (SPT) phase of 1D quantum spin systems is the Haldane phase, which is the ground state phase of the standard antiferromagnetic Heisenberg model with $S=1$. The gaped nature of the ground state phase and the existence of the edge states were predicted a long time ago \cite{Haldane19831, Haldane19832, AKLT}, while no local order was identified to characterize the phase. Later on, Den Nijs and Rommelse \cite{DenNijs} have shown that the Haldane phase has a hidden Neel order, which can be identified by the non-local string order parameter (SOP). Kennedy and Tasaki \cite{kennedy1992hidden} explained the origin of the gap in terms of the hidden $\mathbb{Z}_2 \times \mathbb{Z}_2$  symmetry breaking using non-local unitary transformation. This transformation usually converts SOPs to a simple ferromagnetic order parameter shedding light on the symmetry structure which is protecting the Haldane phase. As a result, for a long time the only condition for existence of the Haldane phase in spin systems was a non-zero value of SOP. Employing this, in the early 1990's the Haldane phase was detected in modified Heisenberg spin chains with $S=1$ and Heisenberg ladders with $S=\frac{1}{2}$ both numerically and analytically \cite{Hida1991, Kennedy1990}. In particular, the Haldane phase of the bond-alternating Heisenberg model was also studied numerically using SOP via exact diagonalization methods \cite{Hida19921, Hida19922}.  

 While the Haldane phase was identified via SOPs in spin systems, in fermionic and bosonic systems rigorous and generalized tools such as the Berry phase have been used. As a result, the question of generalization of fermionic order parameters for spin systems has arisen. Hastugai \textit{et al.}  \cite{Hatsugai} introduced a local spin twist as a generic parameter of Berry phase for gapped spin systems, which has been used extensively \cite{Hatsugai2}. Along this line of thought, Rosch and Anfuso \cite{Anfuso} considered the reverse process, i.e using SOP as an SPT order parameter for fermionic systems. They reconstructed SOP for identification of the SPT phase in band insulators and studied the robustness of the SOP against perturbation terms  \cite{Anfuso2}. More recently, SOP was used to identify the Haldane phase in a 1D bosonic lattice, a topological Kondo insulator and a Kitaev Ladder \cite{KitaevLadder,BosonicSPT,KondoIns}, demonstrating their solid application for non-spin systems. However, these works were performed numerically using DMRG \cite{White1992,Schollwock2011} and SOPs were not evaluated analytically. As the first result of the present paper, we analytically evaluate SOP to identify the SPT phase in the example of a dimerized fermionic model. The current well-studied model is considered due to it's simplicity and it can demonstrate clearly the physical picture of SOPs. This result, particularly, expands the list of exactly calculable order parameters of SPT phases.  
 
 Another question which we address in this letter is the behaviour of local two-site entanglement in the vicinity of topological quantum phase transitions (tQPT). There have been numerous studies carried out on the characterization of topological phases and their transitions in which entanglement entropy and entanglement spectrum are introduced to identify tQPT \cite{entropy, entspectrum1,Pollmann2010} and the corresponding phases. However, only a few studies were carried on the behaviour of local pairwise entanglement in tQPTs \cite{CreutzLadder}. A good measure of local bipartite entanglement is concurrence, defined by Wootters \cite{Wooters}, and it has been used extensively to study trivial QPT occuring in 1D quantum spin chains \cite{Glaser, Discord, Conc1,Osborne,Vidal,Osterloh}.  In particular, in \cite{ConcEnergy} it was shown, that in trivial QPT non-analytic behaviour of the concurrence or its derivative determine the order of the quantum transition. Hence, an interesting question to investigate is the behaviour of microscopic correlations and the concurrence in the vicinity of tQPT, which we discussed in the second part of the paper.

 The paper is organized as follows. In Sec. \ref{sec:Modeldef} we introduce the considered spin model and describe basic steps of fermionization and exact analytical diagonalization of it. Fermionized versions of the SOPs, their exact evaluation and the final phase diagram  are presented in Sec. \ref{sec:FSOPS}. We discuss the behaviour of local bipartite entanglement in the vicinity of topological QPT in Sec. \ref{sec:EntStruct}. Finally, conclusive notes and future outlooks are presented in the last section Sec. \ref{sec:conclusion}.

\section{Model definition}
\label{sec:Modeldef}
We consider a generalized spin-1/2 $XX$ chain of length $N$ with bond-alternation defined as:
 \begin{equation}
  H_1 =J \sum_{i=1}^{N }( S^{y}_{2i-1} S^{y}_{2i}+S^{x}_{2i-1} S^{x}_{2i})   %+ \sin(\theta)(\vec{S }_{2i} \vec{S }_{2i+1} (\Delta^{\prime})  
 \label{Hamiltonian1}
\end{equation}
 \begin{equation}
  H_2 = J^\prime \sum_{i=1}^{N }(S^{y}_{2i} S^{y}_{2i+1}+S^{x}_{2i} S^{x}_{2i+1})    
 \label{Hamiltonian2}
\end{equation}
where $J$,$J^\prime$ are spin couplings.   The sum of these two terms defines the Hamiltionian of our model:
\begin{equation}
 H=H_1+H_2  
 \label{Hamiltonian3}
\end{equation}
 
We note that $J, J^\prime$ can be ferromagnetic (FM) and antiferromagnetic (AFM) . Since the relative absolute value of couplings define ground state phases of (\ref{Hamiltonian3}), we parameterize $J=2\cos(\theta)$ and $J^\prime=2 \sin(\theta)$, where $\theta \in [0, 2\pi)$. 

	Due to broken translational symmetry, one can relabel spins within the effective unit cell and rewrite Eqs.(\ref{Hamiltonian1}-\ref{Hamiltonian2}) in the following form:
	\begin{equation}
	H_{1}=  \cos(\theta) \sum_{i=1}^{\frac{N}{2} }   (  S^{+(a)}_{i}S^{-(b)}_{i}+ S^{-(a)}_{i}S^{+(b)}_{i})   
	\label{Hams1}
	\end{equation}
\begin{equation}
	H_{2}= \sin(\theta)   \sum_{i=1}^{\frac{N}{2}-1 }   (  S^{+(b)}_{i}S^{-(a)}_{i+1}+  S^{-(b)}_{i}S^{+(a)}_{i+1})
	\label{Hams2}
	\end{equation}
where (a) and (b) refer to spins located within the $i$-th unit cell.
For $\theta= \frac{n \pi}{2}$ , $n \in \mathbb Z$ one has localized spin dimers along the chain .  Hereafter, we assume periodic boundary conditions (PBC).

To fermionize (\ref{Hams1}-\ref{Hams2}) we use the Jordan - Wigner transformation \cite{JWT} which we define as:
\begin{equation}
S^{+(a)}_{i}=a^{\dagger}_{i} e^{i \pi \sum_{m<i} (a^{\dagger}_{m} a_{m}+b^{\dagger}_{m} b_{m})}
\end{equation} 
\begin{equation}
S^{+(b)}_{i}=b^{\dagger}_{i} e^{i \pi \sum_{k<i} (a^{\dagger}_{k} a_{k}+b^{\dagger}_{k} b_{k}+a^{\dagger}_{i} a_{i})}
\end{equation} 
\begin{equation}
 S^{z (a)}_{i}=   a^{\dagger}_{i} a_{i} -1/2
\end{equation} 
where operators $ a_{i},b_{i}$ are spinless fermionic operators with standard anticommutation rules. This transformation maps (\ref{Hamiltonian3}) to the following fermionic model:
\begin{equation}
{\cal H}=\sum_{i=1}^{N_c } \cos(\theta) (a^{\dagger}_{i}b_{i} +b^{\dagger}_{i} a_{i}) +\sin(\theta) (b^{\dagger}_{i}a_{i+1}+a^{\dagger}_{i+1}b_i )
\label{HamGen}
\end{equation}
where $N_c$ is the number of effective unit cells. In derivation of (\ref{HamGen}) we neglected boundary terms, assuming that the chain is  long enough, so that the boundary terms' contribution to the physical quantities $ \sim O(\frac{1}{N})$ is negligible. 

 The Hamiltonian (\ref{HamGen}) is a generalized non-interacting Su-Schrieffer-Heeger (SSH) model. It should be noted, that in contrast to the well-studied standard SSH model \cite{SSHmodel,Asboth}, here one may have different signs of the fermionic hopping parameters (due to the freedom of the signs for spin couplings). Furthermore, if the isotropic Dzyaloshinskiy-Moriya  interaction along the chain $H_{DM}=D^z \sum_{i=1}^{L } ( S^{x}_{i} S^{y}_{i+1}-S^{y}_{i} S^{x}_{i+1}) $ is introduced to the model (\ref{Hamiltonian3}), one may have complex values of the hoping couplings $J , J^{\prime} $. This representation freedom of spin chains in terms of fermions enriches the physical phenomena and may correspond to fictitious fermionic models.
    
To express (\ref{HamGen}) in $k$-space, we use the discrete Fourier transformation of the form:
\begin{eqnarray}
a_{j}=\frac{1}{\sqrt{N_c}}\sum_{k \in BZ}e^{-ik j}a_{k},\text{ \ \ }\\b_{j}=
\frac{1}{\sqrt{N_c}}\sum_{k \in BZ}e^{-ik j}b_{k},
\end{eqnarray}

 Then, the Hamiltonian (\ref{HamGen}) in the momentum space can be written as:
\begin{eqnarray}
\cal{H} &=&  \sum_{k \in BZ}
\Gamma_k^{\dagger}
\hat{F}_k
\Gamma_k, \label{FTf }
\end{eqnarray}
where
\begin{equation}
 \hat{F}_k=  
 \begin{bmatrix}
     0&    e^{-i\frac{k}{2}}\sin(\theta)+e^{i\frac{k}{2}}\cos(\theta)    \\
     \sin(\theta)e^{i\frac{k}{2}}+ e^{-i\frac{k}{2}}\cos(\theta)  & 0   \\
\end{bmatrix}
 \end{equation}
and  $\Gamma_k^{\dagger} =(a_{k}^{\dagger},b_k^{\dagger} )$.  
 \begin{figure}
 \includegraphics[width=8cm]{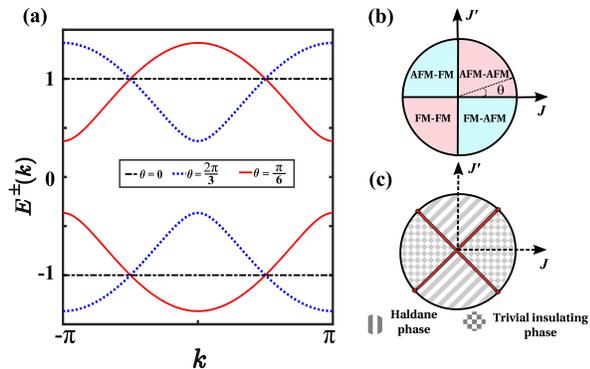}
 \caption{ (color online)  (a) Quasiparticle energy spectra $E^{\pm} (k,\theta)$ for the values of $\theta = [ 0,\frac{\pi}{6} , \frac{2 \pi}{3} ]$. Flat band corresponds to the dimers localized within the unit cell. Remaining dispersive bands represent two modes of the model, as shown in (b). In (c) the phase diagram of the model is shown.  }
  \label{FIG2s}
 \end{figure}
One may get spectrum of the Bogolyubov quasiparticles by diagonalization of the kernel (\ref{FTf }),
\begin{eqnarray}
\cal{H} &=&  \sum_{k \in BZ}
\Psi_k^{\dagger}
\hat{\Lambda}_k
\Psi_k, \label{FTfDiag }
\end{eqnarray}
 with $\Psi^{\dagger}_k = (\alpha_{k}^{\dagger},\beta^{\dagger}_k )$. The matrix elements of the new diagonal kernel $\Lambda_k$ define the spectrum of the quasiparticles:
   \begin{equation}
 E^{\pm} (k,\theta) =  \pm \sqrt{ 1 + \sin(2 \theta) \cos(k)  }  
 \end{equation}
 The vector $\Psi_k$ of qusiparticle operators is connected to the vector $\Gamma_k $ via, 
 \begin{equation}
 \Psi_k= \hat{U}_k \Gamma_k 
 \label{Quasivect}
 \end{equation}
 where $\hat{U}_k$ is the transformation matrix formed out by the eigenvectors, 
 \begin{equation}
 v^{\pm}=\frac{1}{\sqrt{2}}  \mqty( \pm \frac{E^{+}(k,\theta)}{e^{i\frac{k}{2}}\sin(\theta)+e^{-i\frac{k}{2}}\cos(\theta)}\\  1  ) 
 \label{vect}
 \end{equation}
In Fig.\ref{FIG2s} we plot the spectrum of the quasiparticles for different values of intracell and intercell couplings. For $\theta=0$ the intercell coupling vanishes, thus one has localised fermions within the unit cell. This flattens the quisiparticle bands, making them to have infinite mass. This character in fact is preserved for any $\theta= \frac{n \pi}{2}$ , $n \in \mathbb Z$.

  For other values of the couplings, the bands get dispersed with a finite  density of states (DOS).  Due to the freedom on couplings' signs,  there are two different modes for the system: couplings with the same signs (FM-FM and AFM-AFM, red sectors in Fig.\ref{FIG2s} (b)) and couplings with alternating signs (AFM-FM and FM-AFM, blue sectors in Fig.\ref{FIG2s}(b)).
  
   The first mode (red sectors of the unitary circle) has a particle - hole symmetric excitation spectrum with the minimum gap at $k=\pm \pi $, as shown in Fig.\ref{FIG2s}(a)  (for $\theta=\frac{\pi}{6}$).  The value of the gap at  $k=\pm \pi$ is 
  \begin{equation}
  \Delta_{G1} (k=\pm \pi)= 2\sqrt{1\ - \sin(2\theta) } .
  \label{Gap1}
\end{equation}   
These red sectors correspond to the SSH limit, since both couplings have the same signs. 
    In the second mode, which corresponds to the blue sectors of the unitary circle, the intercell (intracell) coupling is ferromagnetic, while intracell (intercell) is antiferromagnetic. Then minimum value of the gap is reached at $k=0$ and can be determined as,
\begin{equation}
  \Delta_{G2}(k=0)= 2\sqrt{1\ + \sin(2\theta) } 
  \label{Gap2}
\end{equation}

 From the last expressions (\ref{Gap1}-\ref{Gap2}) for the gap value, on can see that there are gap closures which occure for $\theta = \frac{(2n+1)\pi}{4} $, $n \in \mathbb Z$. These gap closures hint for a quantum phase transition, which happens to have topological nature.
  
 \section{Fermionized String Order parameters}
 \label{sec:FSOPS}

Previously, the tQPT of the model in the Heisenberg limit was studied via DMRG using SOP and symmetry fractionalization technique \cite{Dimerized1,Rizakhani}. In this section, we show an exact calculation of SOP employing the fermionization approach for our $XX$ model.

 For the generalized model (\ref{Hamiltonian3}) the SOP can be defined as  \cite{Hida1991},
 \begin{equation}
 O^{S (\gamma)}= \lim_{r\to\infty} O^{\gamma} (r)
 \label{SOP}
\end{equation}
 
\begin{equation}
 O^{\gamma}_{l,m}(r)=-4\expval{S^{\gamma (b)}_{l} e^{i \pi (S^{\gamma (a)}_{l+1} + S^{\gamma (b)}_{l+2}+\dots + S^{\gamma (b)}_{m-1})} S^{\gamma (a)}_{m}}
 \label{LRS}
 \end{equation}
 where $r=\abs{m-l}$ and $\gamma \in [x,y,z]$.
 For convenience, we further call (\ref{LRS}) \textit{Long-Range String} (LRS). From Eq.(\ref{SOP}) it is clear that a long-distant limit of LRS defines SOP.
Here, the prefactor $-4$ is used to normalize, so that in the dimerized topological limit $ \theta=\frac{(2n+1)\pi}{2} $  it yields $O^{\gamma}_{l,m}=1$. 
 
For longitudinal LRS, using $e^{i\pi S^z}=\frac{S^z}{2i}$, Eq.(\ref{SOP}) can be  written as: 
 \begin{equation}
 O^{z}_{l,m}=\expval{ e^{i \pi (S^{z (b)}_{l}+S^{z (a)}_{l+1} + S^{z (b)}_{l+2}+\dots + S^{z (b)}_{m-1} + S^{z (a)}_{m})}}
 \end{equation}
Since the model is isotropic, $O^{x}_{l,m}=O^{y}_{l,m}$, one needs to evaluate only one of them. Using $e^{i \pi S^{y}_i}=S^{+}_i-S^{-}_i$, LRS for the y-component can be expressed as:
 
  \begin{equation}
 O^{x,y}_{l,m}=\expval{(S^{+ (b)}_l-S^{-(b)}_l)(S^{+ (a)}_{l+1}-S^{-(a)}_{l+1}) \dots (S^{+ (a)}_{m}-S^{-(a)}_{m})} 
 \end{equation}

By the use of JWT, we convert LRS for spins to fermionic LRS :
 \begin{equation}
 O^{z}_{l,m}=\expval{ e^{i \pi ( b^{\dagger}_{l}b_{l}+ a^{\dagger}_{l+1}a_{l+1} +\dots +  a^{\dagger}_{m}a_{m})}}
 \label{OzL}
 \end{equation}
 which is obtained using the fact that the number of operators in the exponent is always even.
For the transverse LRS we obtain, 
 \begin{equation}
 O^{x,y}_{l,m}=\expval{(b^{\dagger}_l+b_l)(a^{\dagger}_{l+1}-a_{l+1}) \dots (a^{\dagger}_{m}-a_{m})}
 \label{OxyL}
 \end{equation}
  We introduce the following operators  for a more convenient notation:
 \begin{equation} 
 A_m=(a^{\dagger}_m - a_m) 
 \label{A}
 \end{equation}
 \begin{equation} 
 B_m=(a^{\dagger}_m + a_m) 
  \label{B}
 \end{equation}
  \begin{equation} 
 C_m=(b^{\dagger}_m - b_m) 
  \label{C}
 \end{equation}
   \begin{equation} 
 D_m=(b^{\dagger}_m + b_m) 
  \label{D}
 \end{equation}
 It is straightforward to derive anticommutation relations of the pairs for introduced operators:
 \begin{equation}
 \acomm{A_m}{A_n}=\acomm{C_m}{C_n} =-2\delta_{m,n}
 \end{equation}
  \begin{equation}
 \acomm{B_m}{B_n}=\acomm{D_m}{D_n} =2\delta_{m,n}
 \end{equation}
 where $\delta_{m,n}$ is Kronecker delta function.
While for any other pair $N$ and $M$ from the set above, we have:
\begin{equation}
\acomm{M_m}{N_n}=0
\end{equation}
Then, fermionized LRSs (\ref{OzL}-\ref{OxyL}) can be expressed in terms of the introduced $A-B-C-D$ operators,
 \begin{equation}
 O^{z}_{l,m}=\expval{ D_l C_l B_{l+1} A_{l+1} \dots D_{m-1} C_{m-1}  B_m A_m  }
 \label{OzLL}
 \end{equation}
 \begin{equation}
 O^{x,y}_{l,m}=\expval{ D_l A_{l+1} D_{l+2} A_{l+2} \dots A_m   }
 \label{OxyLL}
 \end{equation}
 \begin{figure}
 \includegraphics[width=8cm]{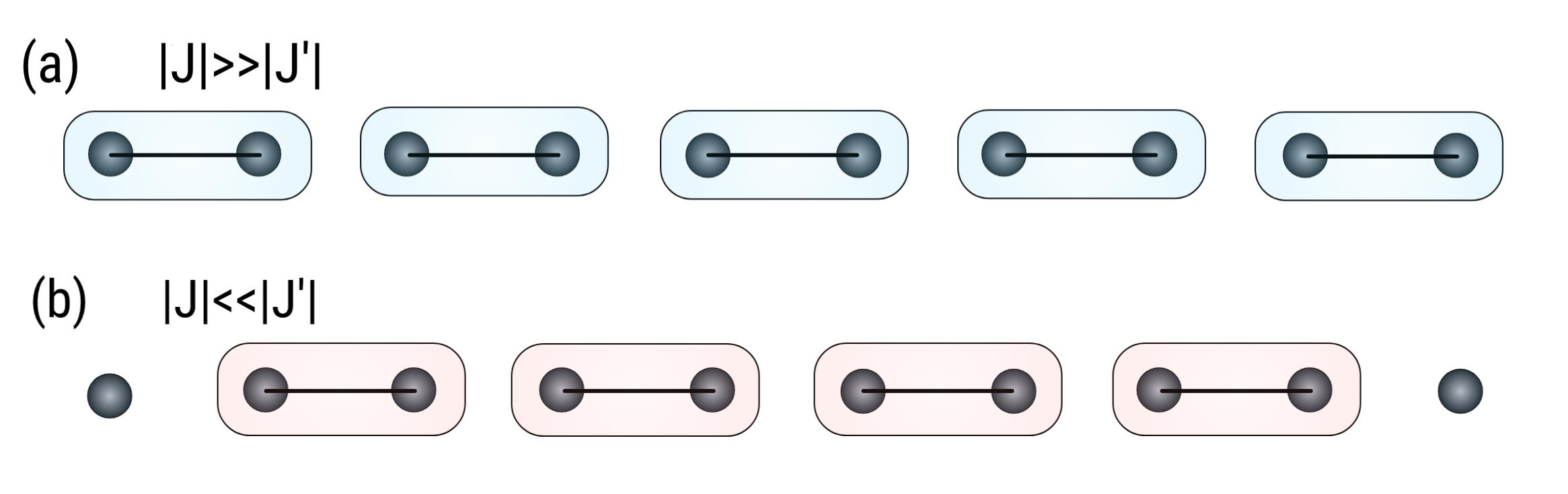}
 \caption{ Schematic representation of the maximally entangled pairs formed in the (a) trivial insulating phase  ( $|J| \gg |J^\prime|$ ) and (b) in the Haldane phase ($|J| \ll |J^\prime|$). For the chain with OBC in the Haldane phase with $|J| \ll |J^\prime|$ one has free spins at the edges, forming corresponding degenerate edge states which are peculiar to the SPT phases. }
  \label{FIG1s}
 \end{figure}
Using Wick theorem, including only non-zero contractions, we get the following result:
 \begin{equation}
O^{z}_{1,m}= \det (K)
\label{Eq35}
 \end{equation}
  \begin{equation}
O^{x,y}_{1,m}= \det (Q)
 \end{equation}
 where by site $l=1 $ we mean any chosen $b$ site as reference and $m$ any chosen site $a$ which follows $b$. Matrix elements $ K_{l,m}$ and $Q_{l,m}$ are defined as, 
 \begin{equation}
 K_{l,m}=\frac{1}{2 \pi} \int_{0}^{2 \pi}   \frac{e^{-ik (m-l) } (e^{-2ik} \cos(\theta)+\sin(\theta))}{ \sqrt{ 1 + \sin(2\theta) \cos(2k)  }} dk 
 \end{equation}
\begin{equation}
 Q_{l,m}=\frac{1}{2 \pi} \int_{0}^{2 \pi}   \frac{e^{-ik (m-l) } (e^{-ik} \cos(\theta)+\sin(\theta))}{ \sqrt{ 1 + \sin(2\theta) \cos(k)  }} dk 
 \end{equation} 
Matrices $K$ and $Q$ are $(m-1) \cross (m-1)$ Toeplitz matrices, with their corresponding
generating functions.
  
Finally, $O^{S (x,y,z)}$ can be evaluated as the determinant of the Toeplitz matrices in the  long range limit,  
\begin{equation}
 O^{S(z)}=\lim_{m \rightarrow \infty} \abs{\det  (K   ) }
\end{equation} 
\begin{equation}
 O^{S(x,y)}=\lim_{m \rightarrow \infty} \abs{\det(Q   )}
 \label{Eq40}
\end{equation} 
 To evaluate SOP $ O^{S(\gamma)} $ analytically, one may use Szeg\"{o} theorem \cite{Szego} in the Haldane phase region of the phase diagram. However, on the other insulating limit this theorem is not applicable, due to the emerging zeros in the corresponding generating functions. 
  
 \begin{figure}
 \includegraphics[width=9cm]{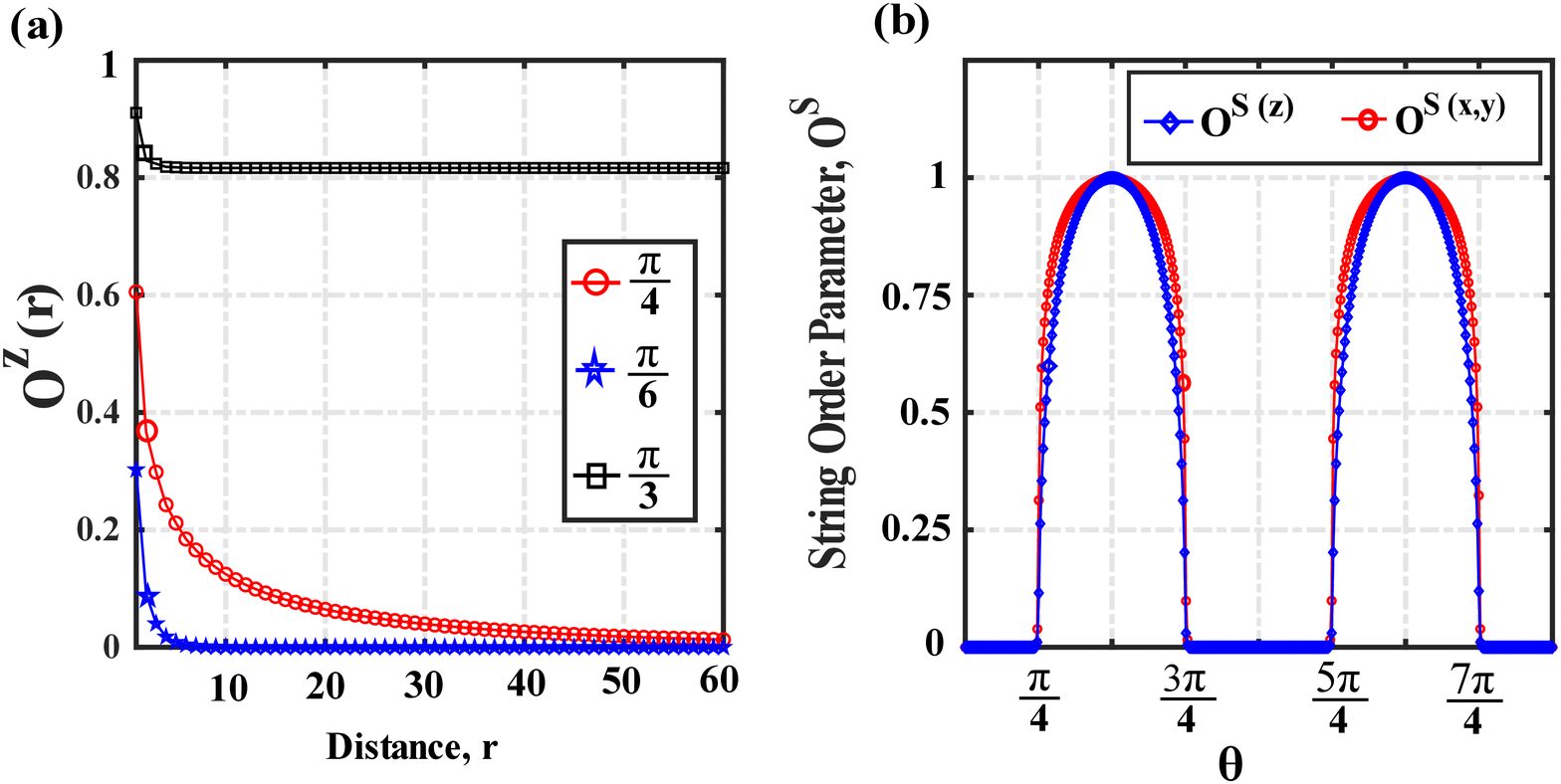}
 \caption{ In (a) LRS $O^{z}(r)$ evaluated in the Haldane phase ($\theta=\frac{\pi}{3}$), in the trivial insulating phase ($\theta=\frac{\pi}{6}$) and in the QCP ($\theta=\frac{\pi}{4}$) are presented. SOPs $O^{S(x,y,z)}$ are evaluated as a long-distant behaviour of Long-range Strings and shown in (b). A region of the unit circle with a finite value of $O^{S(x,z)}$ define the Haldane phase region, which is consistent with the phase diagram represented in Fig.\ref{FIG2s} (c). }
  \label{FIG3s}
 \end{figure}
 It can be easily seen that SOPs are unity when the wavefunction is the tensor product of maximally entangled pairs between the unit cells ($b-a$ pair), i.e 
  \begin{equation}
\ket{\Psi}=\prod_{i=1} ^{ N_c-1} \frac{(\ket{\uparrow^{b}_{i} \downarrow^{a}_{i+1}} \pm \ket{  \downarrow^{b}_{i} \uparrow^{a}_{i+1}} )}{\sqrt{2}}
\end{equation}
This corresponds to the Haldane phase edge limit, i.e $\cot(\theta)=0$. In this limit, free edge spins can be observed in a chain with open boundary conditions as shown in Fig.\ref{FIG1s}, similar to the edge states of the $S=1$ Heisenberg model. However, when the value of the intracell hopping is increased, maximally entangled spin dimers become less entangled, which results in a decreased value of $O^{S(\gamma)}$. Indeed, using Szeg\"{o} theorem one can show that the value of $O^{S(z)}$ for infinitesimal intracell hopping,
\begin{equation}
O^{S(z)} \approx e^{-\frac{\cot(\theta)^2}{4}} ,(    \cot(\theta) \rightarrow 0 )
\end{equation} 
This result shows that for non-zero intracell hopping, $O^{S(z)}$ decreases from unity. The same behaviour can be found for $O^{S(x,y)} $. Thus, $O^{S (\gamma)}$ can be connected to the entanglement measure of $b-a$ spin pairs .  

  In Fig.\ref{FIG3s} we plot the longitudinal LRS $O^z(r)$ for $\theta \in [\frac{\pi}{6}, \frac{\pi}{4}, \frac{\pi}{3}]$ and corresponding SOPs $O^{S(x,z)}$  within the unit circle using Eqs.(\ref{Eq35}-\ref{Eq40}). Fig.\ref{FIG3s}(a) shows that $O^{z}(r)$ exponentially converges to the finite value $O^{S(z)} \approx 0.8$ in the Haldane phase at $\theta=\frac{\pi}{3}$, while exponential convergence to $O^{S(z)}=0$ can be observed in the trivial phase, i.e at $\theta=\frac{\pi}{6}$. In the vicinity of quantum critical points (QCP), $O^{S(z)}(r)$ decreases accroding to a power law, which leads to the divergence of correlation length $\zeta$ at the QCP. Thus, one needs to evaluate $O^{z}(r)$ in the very large distance at these points to extract the value of $O^{S(z)}$. A similar behaviour can be observed on the corresponding phases for $O^{x,y}(r)$. 
 
 In Fig.\ref{FIG3s}(b) we show the phase diagram of the model based on the $O^S$. It demonstrates that for the values of $\abs{\cot(\theta)}<1$, the system is in the Haldane phase, which is characterized by the finite value of SOPs. The maximum values of $O^S$   at $\cot(\theta)=0$. In the QCPs $O^S$ vanish and for the remained circle sectors, i.e $\abs{\tan(\theta)}<1$ the system is in the trivial phase. 
 In analogy to (\ref{LRS}), one may consider the following relation for LRS,
   \begin{equation}
 O^{\gamma (c)}_{l,m}=-4\expval{S^{\gamma (a)}_{l} e^{i \pi (S^{\gamma (b)}_{l+1} + S^{\gamma (a)}_{l+2}+\dots + S^{\gamma (a)}_{m-1})} S^{\gamma (b)}_{m}}
 \label{newSOP}
 \end{equation}
 In contrast to the introduced SOPs, new SOPs derived from Eq.(\ref{newSOP}), detect trivial phase of the model, i.e they are unity when the maximally entangled pairs are $a-b$ spin pairs.   

It should be noted that SOPs are not robust against any symmetry-breaking terms. As an example, an applied magnetic field in $z$ direction which breaks $SU(2)$ symmetry, makes $O^{x,y}_{l,m}$ converge to a finite value even in the trivial insulating phase. An alternative example is the interacting limit, i.e  bond-alternating $XXZ$ chain, for which SOP is a valid order parameter only for the $SU(2)$ symmetric case, i.e in the $XXX$ limit. 
%%%%%%%%%%%%%%%%%%%%%%%%%%%%%%%%%%%%%%%%%%%%%%%%%%%%%%%%%%%%%%%%
\section{Entanglement and Correlations}
\label{sec:EntStruct}

More information about the tQPT discussed in the previous section, can be extracted from a measure of local entanglement. To gain an insight about the entanglement structure of the model, we calculate the two-site density matrix and use two-site concurrence $C^{a,b}_{i,j}$ as an entanglement measure \cite{Wooters}. In trivial QPTs, non-analytic divergent character of the concurrence usually corresponds to first order quantum phase transitions (1QPT), while this behaviour in the derivative of the concurrence is peculiar to the second order transitions (2QPT). Thus, it is interesting to refer to which class of the trivial QPTs the tQPT corresponds. Here, we will exactly evaluate $C^{a,b}_{i,j}$ in terms of two-point correlation functions and investigate the behaviour of it and $\frac{dC^{a,b}_{i,j}}{d\theta}$ in QCPs. 

The functional dependence of the two-site concurrence on standard  correlation functions in generalized quantum spin chains was derived earlier \cite{Glaser} from the reduced density matrix for any two given sites of the chain. Here, we simplify this derivation using symmetries of the Hamiltonian (\ref{Hamiltonian3}). 

The reduced density matrix for any given two sites $i$ and $j$ can be expressed in terms of two-point correlation functions in the following manner (in $\ket{\uparrow \uparrow}$, $\ket{  \uparrow \downarrow}$,$\ket{\downarrow \uparrow}$,$\ket{\downarrow \downarrow  }$ basis):

\begin{equation}
\rho^{(a,b)}_{i,j}=
\begin{bmatrix}
   u^{(a,b)}_{i,j}  & 0 & 0  &0 \\
    0  & w^{(a,b)}_{i,j} & t^{(a,b)}_{i,j} & 0 \\
   0 & t^{(a,b)}_{i,j} & w^{(a,b)}_{i,j} & 0\\
    0 & 0 & 0 & u^{(a,b)}_{i,j} \\
     
\end{bmatrix}  
\label{RhoMatrix}
\end{equation} 
where $u^{(a,b)}_{i,j}=\frac{1}{4}+G^{zz(a,b)}_{i,j}$, $w^{(a,b)}_{i,j}=\frac{1}{4}-G^{zz(a,b)}_{i,j}$ and $t^{(a,b)}_{i,j}=2G^{xx(a,b)}_{i,j}$. Derivation of two-point correlation functions $G^{\gamma \gamma (a,b)}_{i,j}$ for all spin components can be found in Appendix. 

To derive (\ref{RhoMatrix}) we exploited the spin-flip symmetry  (which leads to $\expval{S^{z}}=0$)  and also the symmetry of Hamiltonian (\ref{Hamiltonian3}) with respect to the rotation in the $x-y$ plane. Due to broken inversion symmetry, one has two inequivalent $\rho^{a,b}_{i,j}$. 
 
 \begin{figure}
 \includegraphics[width=9cm]{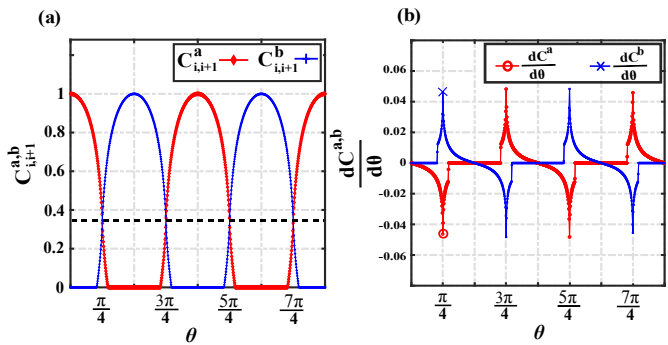}
 \caption{ Calculated pairwise concurrence values for nearest neighbour $a-b$ spin pairs $C^{a}_{i,i+1}$ and nearest neighbour $b-a$ spin pairs $C^{b}_{i,i+1}$ are shown in (a). We note that for the maximally entangled spin pairs $C^{a,b}_{i,i+1}=1$, while decreasing to $C^{a}_{i,i+1}=C^{b}_{i,i+1} \approx 0.35$ (dashed line) at QCP. In these QCPs non-analytic divergent character of the  $\frac{dC^{a,b}}{d\theta}$ can be noticed (b). The points of \textit{sudden death and birth of entanglement } can be noticed with a step like jumps in $\frac{dC^{a,b}}{d\theta}$. }
 \label{FIG4s}
 \end{figure}
 
  The concurrence of two sites $C_{i,j}$ may be computed from the density matrix $\rho_{i,j}$ as $ C_{i,j} = [ \lambda_1 - \lambda_2 - \lambda_3 - \lambda_4 ]$,
where the $\lambda_i$ are the eigenvalues in decreasing order of the matrix $R =\sqrt{\sqrt{\rho } \tilde{\rho}  \sqrt{\rho } }$. The matrix $\tilde{\rho}_{i,j}$  can be expressed as $\tilde{\rho}_{i,j} =(\sigma^{y} \otimes \sigma^{y})\rho^{*}_{i,j} (\sigma^{y} \otimes \sigma^{y})  $. Then, the two site concurrence for any given sites $i$ and $j$  $C^{a,b}_{i,j}$ can be expressed in terms of two-point correlation functions as: 
\begin{equation}
C^{a,b}_{i,j}=2  \text{ max}  \left\lbrace 0, 2 |G^{xx(a,b)}_{i,j}| - |\frac{1}{4}+G^{zz(a,b)}_{i,j}|  \right\rbrace .
\label{concurrence}
\end{equation}

  In Fig.\ref{FIG4s} we present concurrence  $C^a_{i,i+1}$ and $C^b_{i,i+1}$ for nearest-neighbours calculated from Eq.(\ref{concurrence}). For vanishing intracell couplings (i.e at $\theta=\frac{ (2n+1)\pi}{2}, n \in \mathbb Z$), $C^b_{i,i+1}=1$, showing maximal entanglement of the $b-a$ spin pairs. This result is consistent to our SOP result corresponding to maximally entangled $b-a$ spin pairs. Non-zero intracell spin couplings, perturb this entanglement structure of pairs, inducing a decrease in $C^b_{i,i+1}$. This corresponds to the breaking of \textit{entanglement monogamy} of $b-a$ spin pairs. Further decrease of $C^b_{i,i+1}$ continues when the intracell coupling is incremented. However, unlike the SOPs, it doesn't vanish at QCPs. In these gapless Luttinger liquid points every site is entangled with the nearest neighbour with the same concurrence $C^a_{i,i+1}= C^b_{i,i+1} \approx 0.35$ (marked with a dashed line). Further increment of $J$ results in the sudden death of entanglement which happens at $\theta=\frac{ (2n+1)\pi}{2}+\frac{\pi}{3}, n \in \mathbb Z$. 

In the opposite way  $C^a_{i,i+1}$ behaves, which measures an entanglement of $a-b$ spin pairs. Obviously, it is unity for vanishing intercell hopping values $J$. 

As pointed out, the derivative of the concurrence should diverge for  QCPs, at least this is what one  expects in trivial QPTs. Indeed, from Fig.\ref{FIG4s}(b) it is clear that for  $\theta=\frac{ (2n+1)\pi}{4}, n \in \mathbb Z$, one has sharp peaks in $\frac{dC^{a,b}}{d\theta}$ signalling about the tQPT. This character is inherent for the trivial 2QPTs. Similarly, one can show that the second derivative of the ground state energy $\frac{d^2 E}{d \theta^2}$ (exact form of $E$ is presented in Appendix) is also singular at QCPs, which is also a peculiarity of the 2QPTs. Except for these peaks, step like drops of the function can be noticed. These does not effect the phase diagram of the model, however, it is related to the sudden death (or birth) of the entanglement.

\begin{figure}
 \includegraphics[width=9cm]{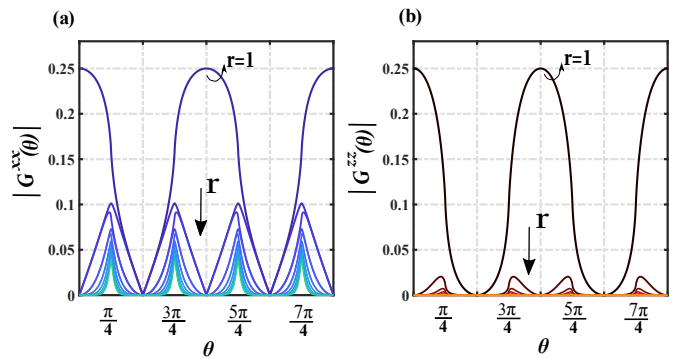}
 \caption{ Modulus of two-point correlation functions $|G^{(a) xx}(\theta)| $ and $|G^{(a) zz}(\theta)|$ of $a-x$ spin pairs are shown in (a) and (b) respectively. For nearest neighbours ($r=1$) the maximum values of absolute correlations are reached in vanishing intercell couplings $J^\prime=0$. For the next nearest neighbours ($r>1$) the maximum values are located at QCPs. Correlation functions $|G^{(a) zz}(\theta)|$ vanish for any $a-a$ spin pairs, i.e for even $r$. }
  \label{FIG5s}
\end{figure}

The concurrences for next neighbours $C^{a,b}_{i,i+r}
=0$ for all values of $\theta$. To show this and to clarify the origin of the singular peaks, in Fig.\ref{FIG5s} we plot the absolute value of the two-point correlation functions $|G^{xx}(r,\theta)|=|\expval{S^{x(a)}_iS^{x(a,b)}_{i+r}}|$ and  $|G^{xx}(r,\theta)|=|\expval{S^{x(a)}_iS^{x(a,b)}_{i+r}}|$ for $ a-x$ pair of spins. To begin with, the derivative of nearest-neighbour correlators have singular peaks at QCP (not shown), resulting in similar peaks of the concurrence. However, derivatives of these correlators for other distances vanish at QCPs, since these points are stationary points. This can be seen from Fig.\ref{FIG5s}(a-b) where the absolute value of correlators for other distances have a maximum value at QCPs. These values decrease with distance, which is known to have a power law behaviour.
One may notice that for $r>1$, the value of $2\abs{G^{xx}}-\abs{1/4+G^{zz}}<0$. Thus, from Eq.(\ref{concurrence}) one concludes that $C^{a,b}_{i,i+r}$ vanish for $(r>1)$, i.e starting from the next nearest neighbours. 
 
In contrast to derived SOPs, two-site concurrences are universal: they identify diverse type of quantum phase transitions even in the presence of symmetry-breaking terms. In this respect, SOPs are limited on application, one should  modify them in a necessary way to determine the phase transition. For this, of course, one needs to have a priliminary knowledge of the corresponding phases. 
 
%%%%%%%%%%%%%%%%%%%%%%%%%%%%%%%%%%%%%%%%%%%%%%%%%%%%%%%%%%%%%%%%
\section{Summary and conclusions}
\label{sec:conclusion}
In this paper, we discussed the implementation algorithm of SOP as an exactly calculable order parameter for the SPT phases and studied the local pairwise entanglement behaviour at tQPT. We mapped the bond-alternating spin chain via Jordan - Wigner fermionization to the generalized version of the SSH model and demonstrated the correspondence of the topological insulating phase of the standard SSH model and the Haldane phase of the bond-alternating spin model. We derived exact analytical expressions for two-point transverse and longitudinal correlation functions. By the fermionization of the longitudinal and transverse SOPs we showed the exact expression of them in terms of the determinant of the Toeplitz matrices with the given generating functions. We demonstrated that the derived expressions of SOPs take their maximal value of unity in the fully dimerized SPT Haldane phase and vanish in the trivial insulating phase. To get more insight about the phase transition, we studied pairwise local entanglement (namely, two-site concurrence) in the vicinity of tQPTs. We found that the two-site concurrence vanishes for all other distances except the nearest neighbour, like in the standard Heisenberg model. The derivative of the concurrence has a singularity at the tQPT, like in trivial second order quantum phase transitions. In contrast to the SOPs, we found the two-site concurrence universal and robust against symmetry-breaking terms.
 
For future outlook, we note that our result of fermionized SOPs can be extended for other 1D spin chains and ladder models. However, the more important topic of fragility of SOPs needs to be investigated in a more detailed way, considering symmetries of the models.  
 
 \begin{acknowledgments} 
M.S.B thanks Prof.Aikaterini Mandilara for useful discussions. 
\end{acknowledgments}
%%%%%%%%%%%%%%%%%%%%%%%%%%%%%%%%%%%%%%%%%%%%%%%%%%%%%%%%%%%%%%%%%%%%%%%%%%%%%%%%%%%%%%%%%%%%%%%%%%%%%%%%%%%%%%%%%%%%%%%%%%%%%%%%
%%%%%%%%%%%%%%%%%%%%%%%%%%%%%%%%%%%%%%%%%%%%%%%%%%%%%%%%%%%%%%%%%%%%%%%%%%%%%%%%%%%%%%%%%%%%%%%%%%%%%%%%%%%%%%%%%%%%%%%%%%%%%%%%

\appendix*
\section{Calculation of zero-temperature correlation functions}
To evaluate the ground states energy per spin, one needs to integrate the band below Fermi energy, which leads to: 
      \begin{equation}
      \epsilon_{G}=\frac{E_G}{N}=-\frac{\abs{\cos(\theta)+\sin(\theta)}}{\pi} E \left( \frac{2\sin(2\theta)}{1+\sin(2\theta)} \right)
      \end{equation}
     where $E$ is the elliptic integral of the second kind. 
	To evaluate static two-point correlator as the function of spin couplings, we rewrite creation and annihilation operators in terms of quasiparticle operators:
	 \begin{equation}
 a_{q}= \frac{1}{\sqrt{2 }A_q }( \beta_q-\alpha_q)
  \end{equation}
  \begin{equation}
 b_{q}= \frac{1}{\sqrt{2}}( \alpha_q+\beta_q )
  \end{equation}
   
  From the definition of quasiparticle vacuum one can get following relationships:
  \begin{equation}
   \expval{b^{\dagger}_k a_q} = -\frac{\delta_{q,k}}{2A_k}
  \end{equation}
 
 \begin{equation}
 \expval{a^{\dagger}_k b_q}=   -\frac{\delta_{q,k}}{2A^{*}_k}
 \end{equation}
   where 
 \begin{equation}
 A_k=\frac{E^{+}(2k)}{e^{i k}\sin(\theta)+e^{-ik} \cos(\theta) }
 \end{equation}
One can show that, 
 \begin{equation}
 \expval{a^{\dagger}_k a_q}=\expval{b^{\dagger}_k b_q}=\frac{\delta_{q,k}}{2}
 \end{equation}
 Using anticommutation relations one can get a remained matrix elements.
 Obviously, for even $r=n-m$ correlation functions  $\expval{a^{\dagger}_ma_n}=\expval{b^{\dagger}_mb_n} $ vanish :
 \begin{equation}
 \expval{a^{\dagger}_ma_n}=\expval{b^{\dagger}_mb_n} = \frac{\sin{(\frac{\pi (n-m)}{2})}}{\pi(n-m)}=0
 \label{vanish}
 \end{equation}  Here, indexes do not correspond to the effective unit cell, but to a site, thus we will work in a \textit{reduced Brillouin zone}. \\
 Next, we evaluate non-vanishing correlation functions $W_{n-m}=\expval{a^\dagger_m b_n}$ and $Z_{n-m}=\expval{b^\dagger_m a_n}$; Further, we assume $n>m$.
 \begin{equation}
  W_{n-m}= -\frac{1}{2\pi} \int_{-\frac{\pi}{2}}^{\frac{\pi}{2}} \frac{e^{-ikr}(e^{ik}\cos(\theta)+e^{-ik}\sin(\theta))}{  \sqrt{ 1 + \sin(2\theta) \cos(2k)  } } dk  
\end{equation}  
  Another  non-vanishing correlator is:
  \begin{equation}
 Z_{n-m}= -\frac{1}{2\pi} \int_{-\frac{\pi}{2}}^{\frac{\pi}{2}} \frac{e^{-ikr}(e^{ik}\sin(\theta)+e^{-ik}\cos(\theta))}{  \sqrt{ 1 +  \sin(2\theta)\cos(2k)   } } dk 
\end{equation} 
It is can be noticed that $W_{r}=Z_{-r}$.
Integrals above for nearest neighbours    (i.e $\abs{ r } = 1$)  are the linear composition of   $K(\sin(\theta),\cos(\theta))$ and $E(\sin(\theta),\cos(\theta)) $ which are elliptic integrals of the first and the second kinds.  For remaining  distances, the analytical solution is involved and the integration consists other terms.
\subsubsection*{ Transverse correlation functions }
We consider correlation function of x - components of spins, i.e $G^{xx(a)}_{m,n}=\expval{S_m^{(a) x} S_n^{(b) x}}$
\begin{equation}
G^{xx(a)}_{m,n}=\frac{1}{4} \expval{(S^{(a) +}_m +S^{(a) -}_m)(S^{(b) +}_n+S^{(b) -}_n}   
\label{Gxx}
\end{equation}
Index $a$ in $G^{xx(a)}_{m,n}$ shows that we look x-component correlation of any $a-b$ spin pair.
In fermionic representation it has the following form: 
 \begin{equation}
G^{xx(a)}_{m,n}=\frac{1}{4} \expval{(a^{\dagger}_m -  a_m) e^{i \pi \sum_{k= m+1}^{k=n-1} (\hat{n}^{(a,b)}_k)} ( b^{\dagger}_n+b_n) }
\end{equation}
Thus, we have for every site one has exponential string operator: 
\begin{equation}
e^{i \pi p^{\dagger}_k p_k}=(1-2p^{\dagger}_k p_k) =(p^{\dagger}_k + p_k)(p^{\dagger}_k -p_k)
\end{equation}
where $p=a,b$ is  fermionic operator.  Equation (A.12) can be written in terms of $A-B-C-D$ operators (\ref{A}-\ref{D}) as: 
 \begin{equation}
G^{xx(a)}_{m,n}=\frac{1}{4} \expval{A_m D_{m+1} C_{m+1} B_{m+2} A_{m+2} \dots  B_{n-1} A_{n-1} D_n  }
\end{equation}
One can evaluate this  correlation function using Wick theorem, since operators anticommute. Thus, we firstly evaluate expectation values of the form $\expval{N_m M_n}$ where $N$ and $M$ belong to a set of operators defined in (\ref{A}-\ref{D}).\\
Using (\ref{vanish}) it can be shown that for different sites  $\expval{N_m N_n}  =0$ for any $N$. Furthermore, $\expval{A_m B_n}  = \expval{C_m D_n} = 0$. 

The only non-zero elements are:
\begin{equation}
\expval{A_mD_n}=2 W_{n-m}
\end{equation}
\begin{equation}
\expval{C_mB_n}=2 Z_{n-m}
\end{equation}
 
At this point, this information is enough to evaluate expression (\ref{Gxx})  using Wick theorem which leads to the following results:
\begin{equation}
G^{xx(a)}_{1,n}=\frac{1}{4} \det 
\begin{bmatrix}
     2W_{1} & 0  &2W_{3} & 0& \dots & 2W_{n-1}  \\  0& 2Z_{1} & 0 & 2Z_{3} & \dots & 0 \\  2W_{-1} & 0  &2W_{1} & 0&\dots & 2W_{n-3} \\  0& 2Z_{-1} & 0 & 2Z_{1} & \ddots & 0 \\ 
 \dots&\dots & \dots &\dots & \dots & \dots 
\end{bmatrix}
\label{GxxG}
\end{equation}

where matrix elements of $G^{xx}$ are defined by $W_{n-m}$ and $Z_{n-m}$. While for the standard XY model \cite{Max} \cite{McCoy} the matrix $G^{xx}$ is a well studied Toeplitz matrix, in our case the matrix is a block Toeplitz matrix, where diagonal elements  oscillate between two values.  \\
We note that using (\ref{GxxG}) $G^{xx(a)}$ can be evaluated for any $a-x$ pair, where $x \in [ a,b] $. Obviously, $G^{xx(b)}_{m,n}  \neq  G^{xx(a)}_{m,n}$. It is straightforward task to show that an equivalent matrix for $G^{xx(b)}_{m,n}$  can be obtained by changing $W_{n-m} \rightarrow Z_{n-m}$. 

Finally, we note that due to the isotropy of the model in $x-y$ plane, the  the correlation function for $y$ - component behaves in the same way.

\subsubsection*{  Longitudinal correlation function }

  Calculation of $G^{zz}_{m,n} $ is not as complex job as transverse correlation functions. We firstly consider $\expval{S^{(a) z}_{m} S^{b (z)}_{n}}$ case, longitudinal correlation of $a-b$ spin pairs:
  \begin{eqnarray}
  G^{zz(a)}_{m,n} = \expval{S^{(a) z}_{m} S^{(b) z}_{n}} = -\frac{1}{4} \expval{ e^{i \pi S^{(a) z}_{m}   } e^{i \pi S^{ (b) z}_{n}  }}  
  \end{eqnarray}
  Fermionized expression is, 
  \begin{equation}
   G^{zz(a)}_{m,n} = -\frac{1}{4} \expval{ e^{i \pi a^{\dagger}_m a_m }   e^{i \pi b^{\dagger}_n b_n  }} 
  \end{equation}
   In terms of $A-B-C-D$ operators,
  \begin{equation}
  G^{zz(a)}_{m,n}=-\frac{1}{4} \expval{  B_m A_m D_n C_n}
 \end{equation}
 The only non-zero contraction is:
 \begin{equation}
 \expval{B_m C_n}\expval{D_nA_m}
 \end{equation}
 Thus, 
 \begin{equation}
 G^{zz(a)}_{m,n}=\frac{1}{4} \expval{B_m C_n}\expval{D_nA_m}=-(W_{n-m})^2
 \end{equation}
Similarly, an expression for a longitudinal correlation function of $b-a$ spin pairs is,
\begin{equation}
 G^{zz(b)}_{m,n} = \expval{S^{(b) z}_{m} S^{(a) z}_{n}}  = - (Z_{n-m})^2
\end{equation}
 For even distances ( $a-a$ spin pair or $b-b$ spin pairs),  $C^{zz(a,b)}_{m,n}$  vanishes, due to the fact that all contractions of the same type of fermions (i.e consisting only $a$ 's or $b$'s) vanish in any permutations. 

%\bibliography{StringSOP.bbl}
% \bibliographystyle{apsrev4-1}

%merlin.mbs apsrev4-1.bst 2010-07-25 4.21a (PWD, AO, DPC) hacked
%Control: key (0)
%Control: author (72) initials jnrlst
%Control: editor formatted (1) identically to author
%Control: production of article title (-1) disabled
%Control: page (0) single
%Control: year (1) truncated
%Control: production of eprint (0) enabled
%

\end{document}